\documentstyle[12pt,aaspp4,epsfig]{article}

\begin{document}                     
\title{Are Extragalactic Gamma Ray Bursts The Source Of The Highest
Energy Cosmic Rays?} 

\author{Arnon Dar}
\affil{Theory Division, CERN, CH-1211 Geneva 23, Switzerland\\  
and\\ 
Department of Physics and
Space Research Institute\\ 
Technion, Israel Institute of Technology, Haifa 32000, Israel} 
 
\begin{abstract} 
Recent observations with the large air shower arrays of ultra high energy
cosmic rays (UHECR) and recent measurements/estimates of the redshifts of
gamma ray bursts (GRBs) seem to rule out extragalactic GRBs as the source
of the cosmic rays that are observed near Earth, including those with the
highest energies.
\end{abstract}

\section{INTRODUCTION}

The origin of high energy cosmic rays (CR), which were first discovered by
V. Hess in 1912, is still a complete mystery (e.g., Berezinskii et al. 
1990; Gaiser 1990, and references therein). Their almost single power-law
spectrum, $dn/dE\sim E^{-\alpha}$, that changes slightly at the so called
`` knee'' around $10^{15.5}~eV$ and at the so called `` ankle'' around
$10^{18.5}~eV$, seem to suggest a single origin of CR at all energies
(Ginzburg 1957; Burbidge 1962; Longair 1981). However, it is generally
believed (e.g, Morrison 1957; Ginzburg and Syrovatskii 1964; Berezinskii et
al. 1990; Gaiser 1990, and references therein) that CR with energy below
the knee are accelerated in Galactic supernova remnants (SNR), those with
energy above the knee may be either Galactic or extragalactic in origin,
and those with energy above the ankle, that are not confined by Galactic
magnetic fields, are extragalactic because of their nearly isotropic sky
distribution (e.g., Takeda et al. 1998; Yoshida and Dai 1997).

\noindent
If the CR accelerators are Galactic, they must replenish for the escape of
CR from the Galaxy in order to sustain the observed Galactic CR intensity. 
Their total luminosity in CR must therefore satisfy, 
\begin{equation}
L_{MW}[CR]=\int\tau^{-1}(Edn/dE)dEdV, 
\end{equation}
where $\tau(E)$ is the mean residence time of CR
with energy $E$ in the Galaxy. It can be estimated from the mean column
density, $X=\int \rho dx$, of gas in the interstellar medium (ISM) that
Galactic CR with energy $E$ have traversed.  From the secondary to
primary abundance ratios of Galactic CR it was inferred 
that (Swordy et al. 1990) 
$X=\bar\rho c\tau\ \approx 6.9 (E/20ZGeV)^{-0.6}~g~cm^{-2}$, 
where $\bar\rho$ is the mean density of interstellar gas along their path.
The mean energy density of CR and the total mass of gas in the Milky Way
(MW), that have been inferred from the diffuse Galactic $\gamma$-ray,
X-ray and radio emissions are, $\epsilon=\int E(dn/dE)dE\sim 1~eV~cm^{-3}$ 
and  $M_{gas}=\int\rho dV\sim \bar\rho V \sim 4.8\times 10^9M_\odot$,
respectively. Hence, simple integration yields (e.g., Drury et al. 1989)    
\begin{equation}
L_{MW}[CR]\sim cM_{gas}\int {Edn/dE\over X}dE 
\sim 1.5\times  10^{41}~erg~s^{-1}.
\end{equation}
The only known Galactic sources which can supply the bulk of the Galactic
CR luminosity are supernova explosions (SNe) (e.g., Ginzburg and
Syrovatskii 1964; V\"olk 1997) and perhaps Galactic
gamma ray bursts (GGRBs) (Dar et al. 1992; Dar et al. 1998), but not
extragalactic GRBs. For completeness and for later
use, we shall first rederive  this result and than proceed to
show that recent data from the large air shower arrays (e.g., Hayashida et
al. 1996; Yoshida and Dai 1998, and references therein) on ultra
high energy cosmic rays (UHECR) and the recent redshift
measurements/estimates of some GRBs and host galaxies of GRBs
(Metzger et al. 1997; Kulkarni et al. 1998; Bloom et al. 1998; Djorgovski
et al. 1998; Fruchter et al. 1998a, 1998b) seem to
rule out extragalactic GRBs as the source of the UHECR.

\section{ARE SUPERNOVA REMNANTS THE MAIN COSMIC RAY SOURCE ?}
Approximately, $E_K\sim 10^{51}~erg$ is released by SNe as nonrelativistic
kinetic energy of ejecta at a rate (Woosley and Weaver 1986), 
$R_{MW}[SNe]\sim 2.5\times 10^{-2}
y^{-1}$. If a fraction $\eta \sim 20\%$ of this energy is converted into
CR energy by collisionless shocks in the supernova remnants (SNR), then
the total SNe luminosity in CR is,
\begin{equation}
L_{MW}[CR]\approx  1.5 \left ( {\eta\over 0.2}\right )
\left ({R_{MW}[SNe]\over 0.025y^{-1}}\right)\left({E_K[SNe]\over
10^{51}erg}\right) \times
10^{41}~erg~s^{-1}, 
\end{equation}
as required by eq.2. Supernova remnants are also natural sites for
Fermi acceleration of cosmic rays by collisionless magnetic shocks and the
SNR environment seems also to explain the chemical composition of CR at
low energies (see, e.g., Ramaty et al. 1998 and
references therein) where it is well measured. Moreover the non thermal
X-ray emission from SNR 1006 observed by ASCA (Koyama et al. 1995) and by
ROSAT (Willingale et al. 1996), the GeV $\gamma$-ray emission from several
nearby SNRs observed by EGRET (Esposito et al. 1996), and the recent
detection of SNR 1006 in TeV $\gamma$-rays by the CANGAROO telescope
(Tanimori et al. 1998), were all interpreted as supportive evidence for
the assumption that SNRs are the source of the bulk of CR. However, the
TeV $\gamma$ rays from SNRs can be explained by inverse Compton scattering
of microwave background photons by multi-TeV electrons whose synchrotron
emission explains their hard lineless X-ray radiation. Furthermore, the
mean lifetime of strong shocks in SNRs limits the acceleration of CR
nuclei in SNRs to energies less than $\sim Z\times 0.1PeV$ (e.g., Lagage
and Cesarsky 1983) and cannot explain the origin of CR with much higher
energies. In fact, the most nearby SNRs in the northern hemisphere have
not been detected in TeV $\gamma$-rays (Buckley et al. 1998).  Moreover,
the scale height of the Galactic distribution of SNRs ($\sim 4.8~kpc$) 
differs significantly from that required $(\geq 20~kpc)$ to explain the
observed Galactic emission of high energy $(>100MeV)$ $\gamma$-rays by
cosmic ray interactions in the Galactic ISM (Strong and Moskalenko 1998). 
Furthermore, the diffusive propagation of CR from the observed/inferred
distribution of Galactic SNRs yields anisotropies that at an energy of
about $100~TeV$ are in excess of the observed value by more than an order
of magnitude (Ptuskin et al. 1997). All these suggest that, perhaps, SNRs
are not the main source of Galacic CR ? 

\section{COSMIC RAYS FROM GGRBs}
Gamma ray bursts (GRBs) have also been proposed as CR sources
(Dar et al. 1992, Waxman 1995, Vietri 1995, Milgrom and Usov 1995;
1996; Dar et al. 1998, Dar 1998, Dar and Plaga 1998). But, if GRBs emit
similar
energies in CR and in $\gamma$-rays (Waxman 1995, Vietri 1995, Milgrom and
Usov 1996), i.e., if  $\Delta E_{CR}\sim \Delta E_\gamma $, then
Galactic GRBs cannot produce the bulk of the CR . This is because
the total CR luminosity due to Galactic GRBs is only,
\begin {equation}
L_{MW}[CR]\sim R_{MW} {4\pi\over \Delta\Omega} E_\gamma {\Delta\Omega\over
4\pi} = 3 \left( {R_G\over 10^{-8}~y^{-1}}\right)\left ({ E_{isot}\over
10^{52}~ erg}\right )\times 10^{36}~erg~s^{-1}, 
\end{equation} 
independent of the solid angle $\Delta\Omega$ which the
gamma ray emission is  beamed into.
The ``isotropic'' energy emission in eq.4 is defined as $E_{isot}\equiv
4\pi(\Delta E_\gamma/\Delta \Omega)$ and  $R_{MW}$  is the rate of 
observable GGRBs (those GRBs in the Milky Way galaxy that emit 
$\gamma$-rays in our direction). 
Wijers et al. (1997) pointed out that if the origin of GRBs is 
related to the birth of neutron stars and black holes, then the
GRB rate is proportional to the star formation rate. In fact, the recent
spectral observations of CGRBs afterglows strongly suggest that GRBs are 
produced in star burst regions. Wijers et al. (1997) used 
the new  distance scale of CGRBs, which follows from the
measured/estimated redshifts of CGRB afterglows and their host galaxies, 
and the assumption that the CGRB rate follows the star formation rate, to
show that the current GRB rate per galaxy is $<2\times 10^{-8}~y^{-1}$.
This
value is two orders of magnitude smaller than that  was thought before.
We have reestimated the current GGRB rate (GRBs in a Milky Way) from new
measurements (Steidel et al. 1998 and references therein)
of the star formation rate as function of redshift $z$, as shown in Fig.1,
using  $R[CGRB]\simeq 10^3~y^{-1}$
for the  rate of observable CGRBs (Fishman and Meegan 1995).
The present $(z=0)$
rate  of observable GGRBs is given approximately by
\begin{equation}
R_{MW}[GRB] \simeq {R[CGRB]L_{MW}R_{SFR}(z=0)\over
               \rho_L \int (1+z)^{-1}R_{SFR}(z)(dVc/dz)dz},
\end{equation}
where $L_{MW}\sim 2.3\times 10^{10}L_\odot$ is the stellar luminosity of
the Milky Way and $\rho_L\simeq 1.8h\times 10^8L_\odot~Mpc^{-3}$ is the
luminosity density in the local universe (Loveday et al. 1992).  For a
critical universe, with ${\Omega}_M=1$ and $\Lambda=0$, one has
$dV_c=16\pi (c/H)^3(1+z-\sqrt{1+z})^2(1+z)^{-7/2}dz$, and the volume
average of the observed star formation rate (Fig.1) yields a mean
rate which is about 15 time larger than that in the local Universe, $\bar
R_{SFR}=\int (R_{SFR}/(1+z)(dV_c/dz)dz/V_c \sim 15R_{SFR}(z=0)$
(the factor 1/(1+z) in the volume integral is the cosmological time
dilation factor). Consequently, with $\int
(1+z)^{-1}(dVc/dz)dz=(16\pi/15)(c/H)^3$ and $h\sim 0.5 $, eq.5 yields
$R_{MW}[GGRB]\sim 2\times 10^{-8}~y^{-1}$, which
is similar to the value obtained by Wijers et al.  (1997).  (The
dependence on $h$, where $H=100h~km~s^{-1}~Mpc^{-1}$ is the Hubble
constant,
cancels out in eq. 5.  The value $h=0.5$ was chosen for the consistency
with Fig. 1).  The result is not much different (but somewhat smaller) for
other standard cosmological models, such as $\Omega_M\sim 0.3$ and
$\Omega_\Lambda\sim 0.7$ or, $\Omega_M\sim 0.2$ and $\Omega_\Lambda\sim 0$.
Thus, we conclude from eqs. 4-5 that GGRBs with integrated CR luminosities
similar to their integrated $\gamma$-ray luminosities, cannot explain the
Galactic CR luminosity. 

\section{COSMIC RAYS FROM CGRBs}
It was suggested independently by Waxman (1995), by Vietri (1995)
and by Milgrom and Usov (1995) that, perhaps, most of the CR luminosity 
of GRBs is in UHECR, and then, isotropically emitting extragalactic GRBs,
with similar integrated CR and
$\gamma$-ray luminosities, may be the source of the UHECR and, perhaps, 
the source of CR with energy down to the knee  (Usov and Milgrom 1996).
However, the mean attenuation length (lifetime) of CR with energies above
about $ 10^{20}eV$, the  so called ``GZK cutoff''
energy in the intensity of UHECR that was predicted independently
by Greisen (1996) and by  Zatsepin
and Kuz'min (1996)  for extragalactic cosmic rays due to their interaction
with the cosmic background photons, is (e.g., Lee 1987) $D<15~Mpc$  ($\tau
<5\times 10^7~y$), as can be seen from Fig. 2. Cascade protons with
an initial particle  spectrum $dn/dE\sim E^{-\beta}$, increse their mean
distance (life time)  from where protons can reach near 
Earth with final
energy $E$, but  they do not change  the  observed spectral index above
the (red shifted) threshold
energy for ``inverse''  photoproduction because of Feynman scaling. The
enhancement factor is given 
approximately by  $k=1/(1-<x>^{\beta-1})^2\simeq
2\pm 0.4 $, where $<x>\rightarrow 0.5$ is the mean fraction of the
initial momentum retained by  protons in inverse photoproduction, 
which is energy independent because of Feynman scaling,
and $\beta\simeq 2.7\pm 0.2 $ is the observed particle spectral index
of the UHECR above the CR ankle. A uniform distribution of galaxies 
(CGRB sites) around  the Milky Way, with a number density $n$ per unit
volume, produces  CR energy flux (energy per unit area, per $sr$, per unit
time),
\begin{equation}
S\approx R_G\Delta E_{CR}{1\over 4\pi}
\int {4\pi r^2 n e^{-r/kD}\over 4\pi r^2}dr
={nR_G\Delta E_{CR}kD\over 4\pi}.
\end{equation}
Cosmic expansion and evolution can be neglected  for cosmological
distances $kD\ll c/H$.  If the
UHECR are trapped locally by (unknown) strong extragalactic magnetic
fields that surround our Milky Way galaxy, then $D$ in eq.6 must be
replaced by $c\tau(E)$, where $\tau(E)$ is the lifetime of UHECR with
energy $E$ in the trap due to attenuation by radiation fields and/or
escape by diffusion in the magnetic fields. 
The measured luminosity density in the local Universe
is (Loveday 1992), $\rho_L\simeq 1.8h\times 10^8L_\odot~Mpc^{-3}$.  If
$R_G< 10^{-8}~y^{-1}$ per $L_*\simeq 10^{10}L_\odot$ galaxy  and if the
kinetic energy release in UHECR per GRB is, $\Delta E_{CR}=5\epsilon
\times 10^{50}~erg$, where $\epsilon$ is
the mean energy of UHECR in $10^{20}eV$ units
(Waxman 1995, Vietri 1995, Milgrom and Usov 1996),
then eq. 6 yields an energy flux of UHECR,
\begin{equation} 
S\simeq 6\left ({n\over
1.8h\times 10^{-2}Mpc^{-3}}\right ) \left ({R_G\over 10^{-8}y^{-1}}\right
)\left ( {\Delta E_{CR}\over 10^{51} erg}\right )  \left ( {kD\over
30Mpc}\right )~eV~m^{-2}s^{-1}sr^{-1}. 
\end {equation} 
The CR above the ankle have an approximate power-law spectrum  (Takeda et
al. 1998), $dn/dE\approx E^{-\beta}$, with $\beta\simeq 2.7\pm 0.$~.  Even
if the bulk of the GRB energy is carried by CR with energy above
$E_0\simeq 10^{20}eV$, one obtains from eq.7, for $E\simeq E_0$, that
\begin{equation} 
E^3{dn\over dE}\simeq (\beta-2)SE_0\left ( {E\over
E_0}\right )^{1-\beta} \sim 4\times 10^{20}eV^2m^{-2}s^{-1}sr^{-1}. 
\end{equation} 
This value is smaller by four orders of magnitude than the observed value
(e.g., Takeda et al. 1998), $E^3dn/dE\simeq 5\times
10^{24}eV^2m^{-2}s^{-1}sr^{-1}$ around $E\sim 10^{20}~eV$. 

\section{COSMIC  RAYS FROM BEAMED CGRBs ?}
\noindent 
The above luminosity problems can be solved by postulating that GRBs emit
isotropically more than $10^{55}$ erg in UHECR, which is very
unlikely for compact stellar objects, or by
jetting the GRB ejecta (Dar et al. 1998; Dar 1998; Dar and Plaga 1998). 
Note that $\Delta E$, the  kinetic energy 
release in GRBs, if they are associated with the
birth of compact stellar objects,  is bounded by their gravitational
binding, and probably it is one or two orders of magnitude smaller,
because of neutrino and gravitational wave emission, 
as observed, for instance, in SNe. But, the ``isotropic''  energy
emission, which is inferred
from the measured  $\gamma$-ray  fluence  $F_\gamma$ of GRBs
their measured redshift $z$ and their  luminosity
distance $d_L$,
\begin{equation}
E_{isot}\equiv 4\pi \Delta E_\gamma/\Delta\Omega\simeq
4\pi d_L^2F_\gamma/(1+z),
\end{equation} 
can exceed even $M_\odot c^2\simeq 1.8\times 10^{54}~erg$ by a large
factor. E.g., if the GRB ejecta is narrowly collimated into a
jet (plasmoid) with a bulk motion Lorentz factor $\Gamma\sim 10^3 $ 
(e.g., Dar and Plaga 1998) then
its radiation is beamed into a solid angle $\Delta\Omega\sim
\pi/\Gamma^2$ and
$E_{isot}\equiv 4\pi
(\Delta E/\Delta \Omega)= (4\pi/\Delta \Omega)\Delta E \sim 4\times 10^6
(\Gamma/10^3)^2 \Delta E$, is much larger than $\Delta E $, the true
energy release in GRBs. Thus, while the total luminosity 
of CGRBs in $\gamma$-rays  (eq.4) 
is independent of the unknown beaming angle, 
$E_{isot}[CR]$ can be much larger than that assumed by Waxman (1995),
Vietri (1995) and  Usov and Milgrom (1996).
However, extragalactic UHECR must show  the GZK ``cutoff'' unless
there is a ``cosmic conspiracy''.  Namely, either the large scale local
magnetic
fields conspire to trap the extragalactic UHECR at the GZK ``cutoff''
energy for a time which is exactly  equal to their attenuation time
in the background radiation (Sigl et al. 1998), or the GRB source spectrum
below the  GZK ``cutoff'' energy is suppressed by exactly the 
attenuation factor above it or GRBs produce new particles 
with a flux that is fine tuned  to produce a smooth CR spectrum 
at the GZK cutoff. 
Such fine tuned ``cosmic conspiracies''  seem very improbable and
unnatural: Observational limits on extragalactic magnetic fields,
from limits on Faraday rotation of radio waves from distant powerful radio
sources (e.g., Kronberg 1994) and from limits on intergalactic
synchrotron emission, imply
Larmor radius for $4\times 10^{19}eV$ protons in typical extragalactic
magnetic fields, $(B<nG)$, that is much larger than the typical coherence
length $(\lambda <1~Mpc)$ of these fields. Moreover, magnetic trapping is
completely ruled out if the arrival directions of UHECR coincide with the
directions of cosmological GRBs (Usov and Milgrom 1995) or if the arrival
directions of
extragalactic UHECR are clustered (Hayashida et al. 1996). Fermi or
collisionless 
shock acceleration normally produce smooth power-law source spectra
and not ad hoc imposed thresholds. Thus, jetting the ejecta of GRBs may
solve the energy problem but does not seem to explain the absence of the
GZK cutoff. 

\noindent
Moreover, UHECR  have a  Larmor radius, 
$R_L\sim 100 (E/10^{20}eV)(ZB/nG)^{-1}~Mpc$, that  is   
much larger than the coherence length, 
$\lambda\sim 1~Mpc $ and $\lambda\sim 1~kpc $,  
of, respectively, the intergalactic 
and the halo  turbulent magnetic fields (Kronberg 1994).
Therefore, they suffer only small random deflections along their
arrival trajectories from  a typical distances $kD\sim 30 Mpc$. This
implies  that the arrival directions of the UHECR  point back in the
directions of nearby ($kD<50~ Mpc$) galaxies, i.e. in the direction of the
Virgo cluster and the  Super galactic plane, which is not observed (e.g.,
Hayashida et al. 1996). Furthermore,  the spread in their arrival
direction
with respect to the source direction has 
r.m.s. angular deviation,
\begin{equation}  
\Delta \theta\sim 2^0 \left ({E\over 10^{20}}\right )^{-1}
               \left ({kD\over 50Mpc}\right )^{1/2}
               \left ({\lambda\over Mpc}\right )^{1/2}
               \left ({ B\over nG}\right ),
\end{equation} 
and arrival times that are spread  with  r.m.s. value
\begin{equation}  
\Delta t\sim 7\times 10^4 \left ({E\over 10^{20}}\right )^{-2}
               \left ({kD\over 50Mpc}\right )
               \left ({\lambda\over Mpc}\right )
               \left ({ B\over nG}\right )^2 ~y.
\end{equation} 
The number of GRBs within  distance of $d\leq 50~Mpc$ during 
this spread of arrival times is $N_{GRB}\simeq (\rho_L/L_{MW})(4\pi/3)d^3
R_{MW}[GGRB]\Delta t < 1$. Consequently, all UHECR with energy above
$10^{20}~eV$ should point back to one or two  sources with an
angular  spread of $\sim 2^0$, which is inconsistent with their observed
wide sky distribution (Takeda et al. 1998).
 
\section{CONCLUSION} 
The above arguments can be repeated for UHECR nuclei that are
photodissociated by cosmic background photons, and for UHECR photons
that are attenuated by pair production. Both have  attenuation
lengths shorter than that of UHECR. Then, it leads to the conclusion
that if the CR that 
are observed near Earth  are  long lived normal CR particles,
their source is not  extragalactic GRBs. However, if GRBs
are narrowly collimated, then Galactic GRBs can produce the cosmic
rays which are observed near Earth, including those with the highest 
observed energies (Dar and Plaga 1998).

{\bf acknowledgement:} The author would like to thank P. Gondolo, R. Plaga
and L. Stodolsky for valuable discussions and suggestions.

\noindent
\begin{figure}
\plotone{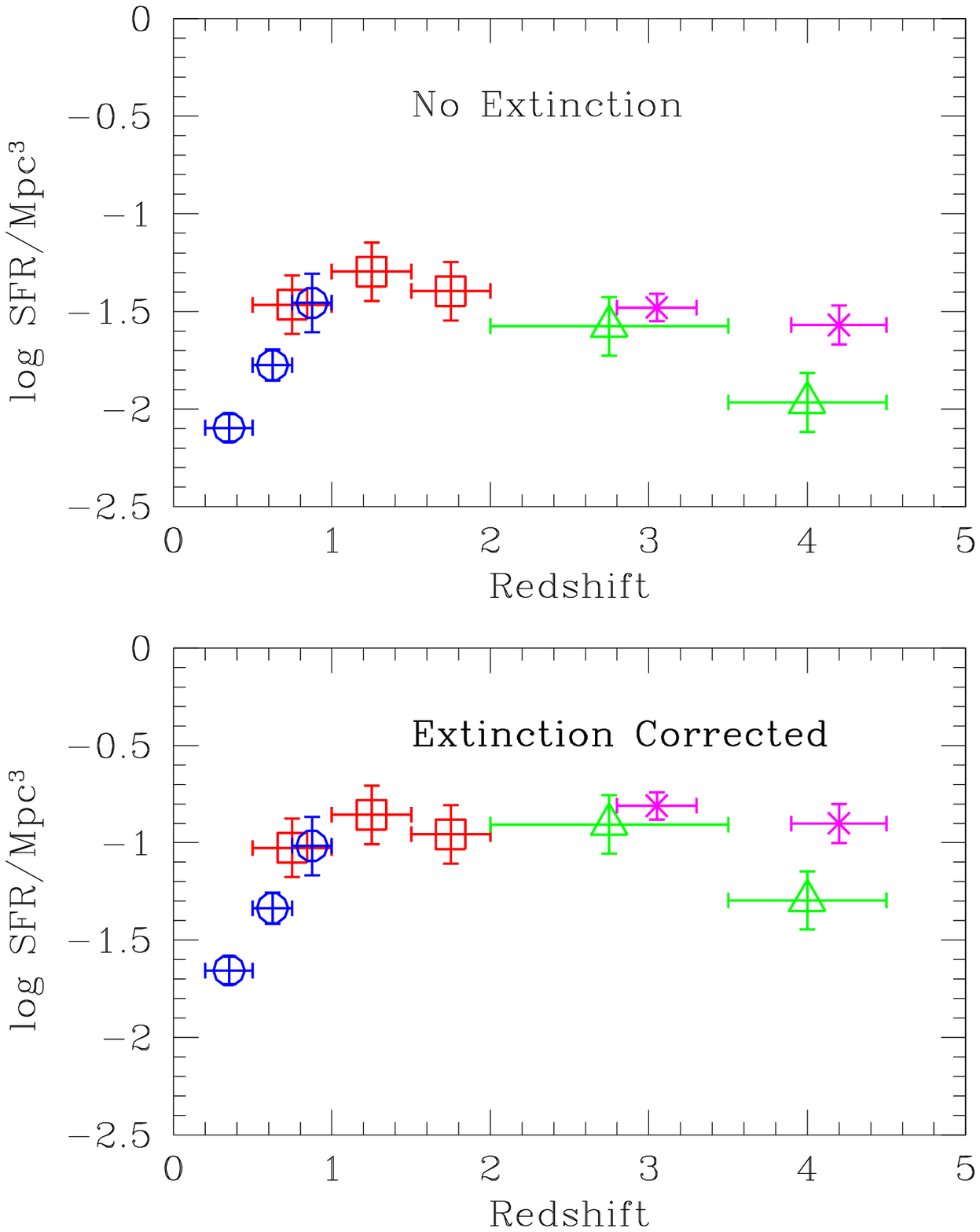}
\caption{The star formation rate per comoving volume as a 
function of redshift, assuming $H_0=50$ kms Mpc$^{-1}$ and $q_0=0.5$,
uncorrected and corrected for extinction by Steidel et al. 1998. The
different points are from Lilly et al.  1996 [circles], Connolly et al.
1997 [squares], Madau et al. 1997 [triangles], and Steidel et al. 1998
[crosses].}
\end{figure}
\newpage
\begin{figure}
\figurenum{2}
\figurenum{2}
\plotone{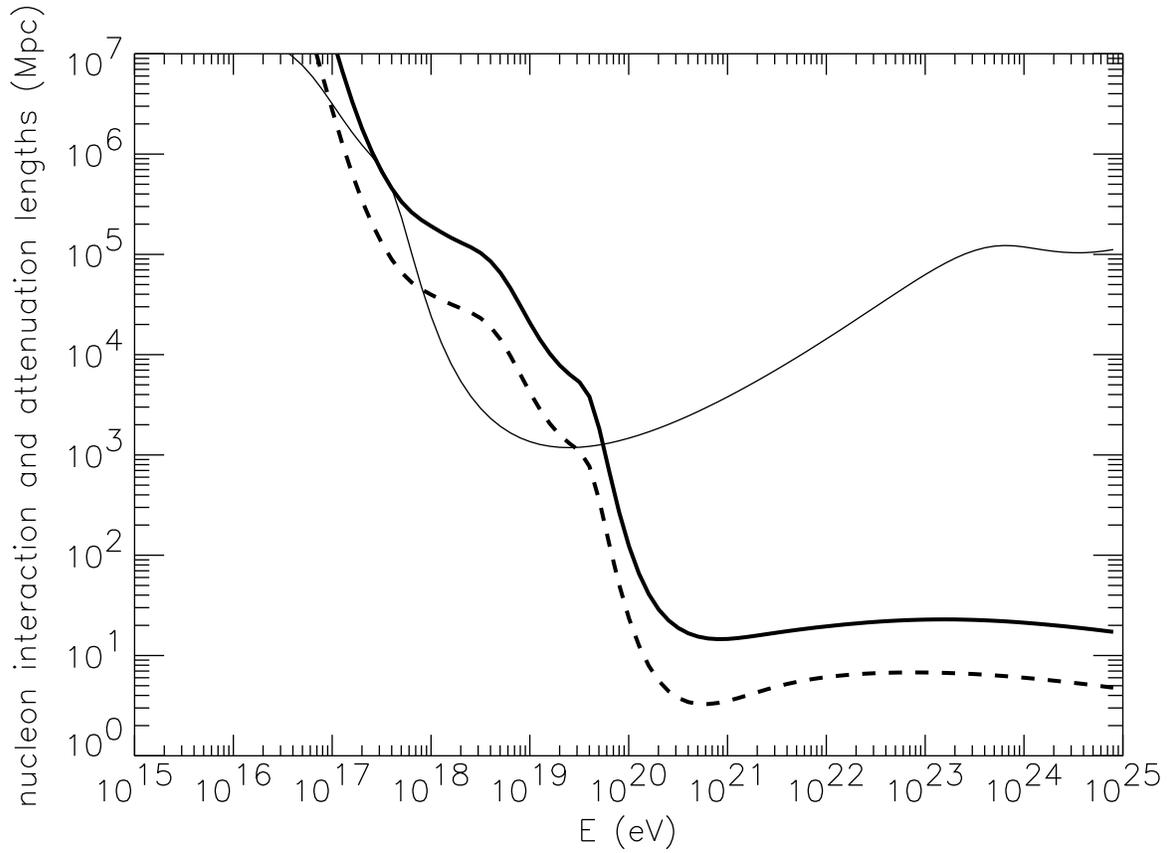}
\caption{The proton interaction length (dashed line) and attenuation
length (heavy line) for inverse photoproduction on the cosmic
background radiation, and  the proton attenuation 
length due to pair production (thin line), as calculated by Lee 1998.}
v\end{figure}

\end{document}